\documentclass[aps]{revtex4}
\usepackage{epsfig}
\usepackage{rotating}
\def\tbar{$\overline{t}~$}               
\def\ppbar{$p\overline{p}~$}             
\def\qqbar{$q\overline{q}~$}             
\def\ttbar{$t\overline{t}~$}             
\def\et{$E_T$}                          
\def\met{\mbox{${\hbox{$E$\kern-0.6em\lower-.1ex\hbox{/}}}_T$}} 
 
\def\D0{D\O}                            
\newcommand{\Dzero}{D\O\ }

\begin{document}
\lefthyphenmin=2
\righthyphenmin=3

%
%
\title{
Search for Pair Production of Light Scalar Top Quarks in \ppbar Collisions at $\sqrt{s}$=1.8 TeV}
\author{
V.M.~Abazov,$^{21}$
B.~Abbott,$^{54}$
A.~Abdesselam,$^{11}$
M.~Abolins,$^{47}$
V.~Abramov,$^{24}$
B.S.~Acharya,$^{17}$
D.L.~Adams,$^{52}$
M.~Adams,$^{34}$
S.N.~Ahmed,$^{20}$
G.D.~Alexeev,$^{21}$
% ...
\\
\vskip 0.30cm
\centerline{(D\O\ Collaboration)}
\vskip 0.30cm
}
\address{
\centerline{$^{1}$Universidad de Buenos Aires, Buenos Aires, Argentina}
\centerline{$^{2}$LAFEX, Centro Brasileiro de Pesquisas F{\'\i}sicas,
                  Rio de Janeiro, Brazil}
\centerline{$^{3}$Universidade do Estado do Rio de Janeiro,
                  Rio de Janeiro, Brazil}
\centerline{$^{4}$Institute of High Energy Physics, Beijing,
                  People's Republic of China}
\centerline{$^{5}$Universidad de los Andes, Bogot\'{a}, Colombia}
\centerline{$^{6}$Charles University, Center for Particle Physics,
                  Prague, Czech Republic}
\centerline{$^{7}$Institute of Physics, Academy of Sciences, Center
                  for Particle Physics, Prague, Czech Republic}
\centerline{$^{8}$Universidad San Francisco de Quito, Quito, Ecuador}
\centerline{$^{9}$Laboratoire de Physique Subatomique et de Cosmologie,
                  IN2P3-CNRS, Universite de Grenoble 1, Grenoble, France}
\centerline{$^{10}$CPPM, IN2P3-CNRS, Universit\'e de la M\'editerran\'ee,
                  Marseille, France}
\centerline{$^{11}$Laboratoire de l'Acc\'el\'erateur Lin\'eaire,
                  IN2P3-CNRS, Orsay, France}
\centerline{$^{12}$LPNHE, Universit\'es Paris VI and VII, IN2P3-CNRS,
                  Paris, France}
\centerline{$^{13}$DAPNIA/Service de Physique des Particules, CEA, Saclay,
                  France}
\centerline{$^{14}$Universit{\"a}t Mainz, Institut f{\"u}r Physik,
                  Mainz, Germany}
\centerline{$^{15}$Panjab University, Chandigarh, India}
\centerline{$^{16}$Delhi University, Delhi, India}
\centerline{$^{17}$Tata Institute of Fundamental Research, Mumbai, India}
\centerline{$^{18}$CINVESTAV, Mexico City, Mexico}
\centerline{$^{19}$FOM-Institute NIKHEF and University of
                  Amsterdam/NIKHEF, Amsterdam, The Netherlands}
\centerline{$^{20}$University of Nijmegen/NIKHEF, Nijmegen, The
                  Netherlands}
\centerline{$^{21}$Joint Institute for Nuclear Research, Dubna, Russia}
\centerline{$^{22}$Institute for Theoretical and Experimental Physics,
                   Moscow, Russia}
\centerline{$^{23}$Moscow State University, Moscow, Russia}
\centerline{$^{24}$Institute for High Energy Physics, Protvino, Russia}
\centerline{$^{25}$Lancaster University, Lancaster, United Kingdom}
\centerline{$^{26}$Imperial College, London, United Kingdom}
\centerline{$^{27}$University of Arizona, Tucson, Arizona 85721}
\centerline{$^{28}$Lawrence Berkeley National Laboratory and University of
                  California, Berkeley, California 94720}
\centerline{$^{29}$California State University, Fresno, California 93740}
\centerline{$^{30}$University of California, Irvine, California 92697}
\centerline{$^{31}$University of California, Riverside, California 92521}
\centerline{$^{32}$Florida State University, Tallahassee, Florida 32306}
\centerline{$^{33}$Fermi National Accelerator Laboratory, Batavia,
                   Illinois 60510}
\centerline{$^{34}$University of Illinois at Chicago, Chicago,
                   Illinois 60607}
\centerline{$^{35}$Northern Illinois University, DeKalb, Illinois 60115}
\centerline{$^{36}$Northwestern University, Evanston, Illinois 60208}
\centerline{$^{37}$Indiana University, Bloomington, Indiana 47405}
\centerline{$^{38}$University of Notre Dame, Notre Dame, Indiana 46556}
\centerline{$^{39}$Iowa State University, Ames, Iowa 50011}
\centerline{$^{40}$University of Kansas, Lawrence, Kansas 66045}
\centerline{$^{41}$Kansas State University, Manhattan, Kansas 66506}
\centerline{$^{42}$Louisiana Tech University, Ruston, Louisiana 71272}
\centerline{$^{43}$University of Maryland, College Park, Maryland 20742}
\centerline{$^{44}$Boston University, Boston, Massachusetts 02215}
\centerline{$^{45}$Northeastern University, Boston, Massachusetts 02115}
\centerline{$^{46}$University of Michigan, Ann Arbor, Michigan 48109}
\centerline{$^{47}$Michigan State University, East Lansing, Michigan 48824}
\centerline{$^{48}$University of Nebraska, Lincoln, Nebraska 68588}
\centerline{$^{49}$Columbia University, New York, New York 10027}
\centerline{$^{50}$University of Rochester, Rochester, New York 14627}
\centerline{$^{51}$State University of New York, Stony Brook,
                   New York 11794}
\centerline{$^{52}$Brookhaven National Laboratory, Upton, New York 11973}
\centerline{$^{53}$Langston University, Langston, Oklahoma 73050}
\centerline{$^{54}$University of Oklahoma, Norman, Oklahoma 73019}
\centerline{$^{55}$Brown University, Providence, Rhode Island 02912}
\centerline{$^{56}$University of Texas, Arlington, Texas 76019}
\centerline{$^{57}$Rice University, Houston, Texas 77005}
\centerline{$^{58}$University of Virginia, Charlottesville, Virginia 22901}
\centerline{$^{59}$University of Washington, Seattle, Washington 98195}
}


\begin{abstract}
 Using $85.2 \pm 3.6$ pb$^{-1}$ 
of $p{\bar p}$ collisions collected at $\sqrt{s}$=1.8 TeV with the \Dzero 
detector at Fermilab's Tevatron Collider, we present the results of a  search for direct pair production of scalar top quarks ($\tilde{t}$),
the supersymmetric partners of the top quark.  We examined events containing two or more jets and 
missing transverse energy, the signature of light scalar top quark decays 
to charm quarks and neutralinos. After selections, we observe 27 events 
while expecting 31.1 $\pm$ 6.4 events from known standard model processes.
Comparing these results to next-to-leading-order production cross 
sections, we exclude a significant region of $\tilde{t}$ and  
neutralino phase space.  In particular, we exclude the $\tilde{t}$ mass $m_{\tilde{t}}<$
122 GeV/$c^2$ for a neutralino mass of 45 GeV/$c^2$.
\end{abstract}
\maketitle

Supersymmetry (SUSY) \cite{SUSY1,SUSY2,SUSY3}, one of the major extensions of the standard model (SM), introduces 
additional particle states. For every bosonic SM 
particle, it assigns a fermionic ``superpartner'' and for every SM fermion, a 
boson. The hypothesized SUSY particles include gauginos 
and scalar quarks or ``squarks.'' The gauginos, superpartners
of the gauge particles, include neutralinos (prime candidates for dark matter). Squarks 
include the left-handed and right-handed scalar top quarks or top squarks.  These weak eigenstates 
mix to provide the mass eigenstates $\tilde{t_1}$ and $\tilde{t_2}$. 

Generic SUSY searches often make the simplifying assumption of mass-degeneracy of
 first and second generation squarks.  The scalar top quark
masses, however, are expected to be substantially smaller than those of 
all other
squarks \cite{stop1,stop2,stop3}.  If sufficiently light, 
scalar top quarks should be produced strongly at the Tevatron through \qqbar 
annihilation and gluon-gluon fusion with a cross section on the order of 
that of the top quark \cite{light_stop,heavy_top}.  
According to the next-to-leading order (NLO) program {\sc prospino} \cite{prospino},  a 100 GeV/$c^2$ scalar top quark has a
 production cross section of about 12 pb, and a 120 GeV/$c^2$ 
scalar top quark of approximately 4.2 pb.  

This analysis is sufficiently general that it applies to a broad class of 
SUSY models.  We make no assumptions about gaugino unification, but assume
 that the lightest neutralino $\tilde{\chi}^0_{1}$ is the lightest supersymmetric 
particle (LSP), with conservation of $R$-parity guaranteeing its stability.  
We consider the special case where the scalar top quark is light enough that 
$m_{\tilde{t_1}} < m_{b} + m_{W}+ m_{\tilde{\chi}^0_{1}}$ and
$m_{\tilde{t_{1}}} < m_{b} + m_{\tilde{\chi}^+_{1}}$, precluding the 
decays $\tilde{t_{1}}\rightarrow b W \tilde{\chi}^0_{1}$,
$\tilde{t_{1}} \rightarrow b \tilde{\chi}^{+*}_{1}$, and  
$\tilde{t_{1}} \rightarrow b \tilde{\chi}^{+*}_{1}$ $(\tilde{\chi}^{+*}_{1} \rightarrow l^+ \tilde{\nu}$ 
or $\tilde{\chi}^{+*}_{1} \rightarrow \tilde{l}^+ \nu )$. The dominant decay is 
then $\tilde{t}_{1} \rightarrow c \tilde{\chi}^0_{1}$, yielding an event 
signature of two jets with 
missing transverse energy (\met). 
We make no attempt to tag the $b$ or $c$ hadrons in jets.  

Characteristics of scalar top quark signal were studied by generating Monte Carlo (MC)
 events for various combinations of $m_{\tilde{t_{1}}}$ and 
$m_{\tilde{\chi}^0_{1}}$,  
using {\sc isajet} \cite{ISAJET} with its implementation of {\sc isasusy}
\cite{ISASUSY}.  These events were processed through a {\sc geant} \cite{GEANT} simulation of the \Dzero detector, a 
simulation of the trigger, and the
standard \Dzero reconstruction program.

The major SM backgrounds expected for 
this signal are multijet events with artificial \met $\:$  and vector 
boson (VB) production associated with associated jets.
The VB backgrounds include those producing: 
neutrinos and jets 
($Z+2$ jets 
$\rightarrow \nu \nu$+2 jets and $W$+jets, where the $W$ boson decays to a 
hadronically-decaying $\tau$ lepton),  
leptons from VB decays that escape 
detection, or electrons misidentified as jets.
{\sc pythia}  \cite{PYTHIA} was used to 
predict the acceptance for $W$/$Z$+jet production, while the {\sc vecbos} 
\cite{VECBOS1,VECBOS2}  Monte Carlo generator was used  for 
$W$/$Z+2$ jets events.  In each case, the calculated cross sections  
were scaled to match internal \Dzero reconstruction and acceptance studies for $W$/$Z$ + $n$ jets.
We also used the cross section for \ttbar 
production measured at \Dzero \cite{ttbar} and the {\sc herwig}
generator to calculate the acceptance for \ttbar background arising from 
top quark decays to an undetected charged lepton, a neutrino, and a jet.

The data correspond to an integrated luminosity of 
85.2 $\pm$ 3.6 pb$^{-1}$ 
collected during the 1994 -- 1995 Tevatron run.  The \Dzero detector
consisted of a central tracking system and a
uranium/liquid-argon calorimeter surrounded by a toroidal muon
spectrometer.  A detailed description of the \Dzero detector and data
collection system can be found in Ref.~\cite{detector}.  
Events were collected using a trigger requiring two jets, one with \et $\;$ $>$ 25 
GeV and 
the second with \et $\;$ $>$ 10
GeV, and \met $>$ 25 GeV, but rejecting events in which
the direction of the leading jet 
and the \met $\:$ are aligned within a polar angle of  $14^{\circ}$.
Jets are reconstructed offline using an iterative cone algorithm \cite{d0jets} of radius 0.5 in
$\eta-\phi$ space.
A requirement of at least 2 jets with
\et $>$50 GeV, \met $>$ 40 GeV, and {\it all jets} satisfying
a difference in azimuth $\Delta \phi(jet,\met)>30^{\circ}$,
guaranteed full trigger efficiency.
To suppress VB backgrounds we removed events with electrons
or muons with \et $>$10 GeV. 

Multijet backgrounds dominate this sample and arise when mismeasured jets or a misidentified
interaction vertex induce an apparent \met. Requiring $\Delta \phi($jet$,\met)<165^{\circ}$ eliminates events 
with jets back-to-back to the \met.  We reduce the number of events with poorly measured jet energies by requiring that the $\Delta \phi$ between the \met$\:$and the jet with the second highest \et$\:$exceed 60$^\circ$. We also removed those events in which jets deposited most of their energy within the narrow intercryostat region ($0.8<|\eta|<1.2$) where the central and endcap calorimeters meet.   We refer to the 354 events surviving these criteria as our
{\em base} sample. (See Table \ref{observed}).

To reduce the background of mismeasured vertices, the central drift 
chamber (CDC) was used to associate charged tracks with jets within the fiducial volume of the CDC, $|\eta|<1$.  Event-by-event these tracks establish
the origin of each jet, which was required to be no further than 8 cm from
the reconstructed event vertex.  This {\em{vertex confirmation}} was 80\% efficient for $W\rightarrow e\nu$ data samples in which electron tracks matched to electromagnetic calorimeter showers provided well-defined interaction vertices, while keeping the mis-matched rate below 2\%.  Table \ref{observed} lists the observed number of events from the jets plus \met $\:$ sample that survive each selection cut down to this {\em clean} sample.

\begin{table}
{\begin{tabular}{cc}
Selection&
Events\\
\hline
\hline
2 jets and \met $\:$trigger&
536,678\\

No detector malfunction or accelerator noise&
487,715\\

Leading jet \et
$>$50 GeV&
205,461\\

Second jet \et
$>$50 GeV&
106,505\\

\met
$>$40 GeV&
13,752\\
$30^\circ < \Delta \phi$(jet, \met) $<165 ^\circ$ & 4650 \\

$60^\circ < \Delta \phi$(jet 2, \met) & 2327\\
Lepton rejection&
2009\\

All jets reside outside \Dzero intercryostat region&
354\\

Vertex confirmation&
88\\
\hline
\end{tabular}\par}
\caption{\label{observed}Number of events surviving in the jets + \met $\:$ sample after the application of selection criteria. 
These 88 events form the {\em clean} sample.}
\end{table}

\begin{table}
{\centering \begin{tabular}{c|c|c}
Source
&  $\:$ Events in {\em clean} sample $\:$ &
$\:$ Events in optimized sample\\
\hline
\hline
W/Z& 63.0 $\pm$ 6.9 $^{+18.1}_{-12.4}$ & 24.2 $\pm$ 3.6 $^{+9.0}_{-6.3}$
\\
$t$\tbar &3.9 $\pm$ 0.02 $^{+0.2}_{-0.5}$ & 3.4 $\pm$ 0.02 $^{+0.2}_{-0.04}$
\\
QCD Multijet & 22.5 $\pm$ 7.5 & 3.6 $\pm$ 1.4
\\
\hline
Total Background & 89.5 $\pm$ 14.7 & 31.1 $\pm$ 6.4
\\
\hline
Data & 88 & 27
\\
\hline
\end{tabular}\par}
\caption{\label{rgsearch_cut_table} A comparison of Standard Model and QCD multijet backgrounds to
the number of candidates in the {\em clean} and final RGS optimized
samples.  For W/Z/$t$\tbar the first uncertainty is statistical, the second
systematic.  For QCD and the total observed, the statistical and systematic
uncertainties have been added in quadrature.}
\end{table}

To predict the multijet background remaining in the {\em clean}  sample, we 
used events from the {\em base} sample where the 
jet vertex position deviated by 15--50 cm from the event vertex.
We normalized this {\em background} sample to the {\em clean} sample 
using events
with $\Delta\phi ($jet$~2,~$\met$) < 60 ^{\circ}$ (where jet 2 refers to the jet with the second largest \et).  
We chose the 50 cm value because it
provides the best agreement between the background prediction and the data
for the \met $\:$ region between 30 and 40 GeV, which is dominated by 
multijet events.  Changing this value to 100 cm (the full width of the 
instrumented interaction region) increases the multijet prediction by 22\%, 
which we take as an estimate of the systematic uncertainty of the method. 
Reversing the order of the vertex confirmation and $\Delta\phi$(jet 2,\met) 
selection with no change in the relative pass rate of events from our {\em 
base} sample showed they provide a legitimate criteria for separating subsets 
for this study. VB background, which include, in decreasing order of 
importance $W\rightarrow\tau + \nu \:+$ 2 jets, $Z\rightarrow\nu+\nu$ + 2 jets and 
$W\rightarrow\mu$ + 2 jets, is comparable to the predicted background 
from multijet production (see table \ref{rgsearch_cut_table}).

\begin{figure}
\begin{minipage}[htb]{8.8cm} 
\epsfysize = 8.5cm 
\begin{turn}{270}
\centerline{\epsffile{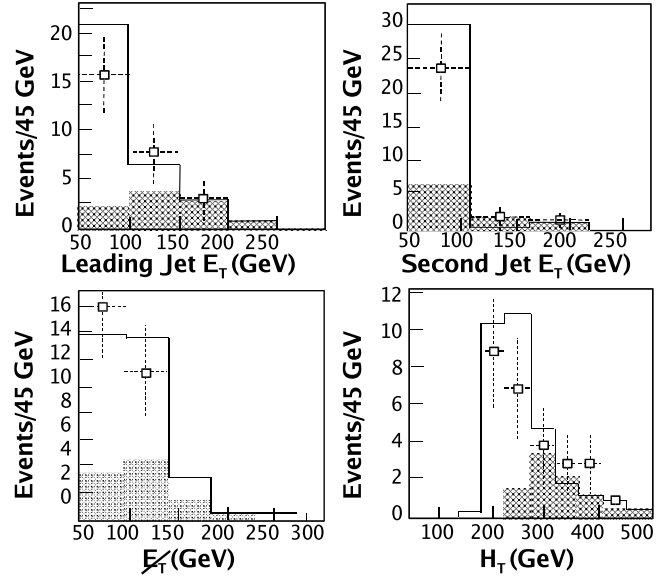}}
\end{turn}
\end{minipage}
\caption{\label{fig:dvsb_final}Data (points) and  predicted background 
(histograms) after final selections.  The additional contributions expected
from a $M_{\tilde{t}}$=130 GeV/$c^2$, $M_{\tilde{\chi_0}}$=30 GeV/$c^2$ scalar top quark are shown by shaded histograms.  
The plots correspond to the \et $\:$ of the leading jet, second jet, \met $\:$ (the three parameters optimized using the RGS) and $H_T$, where 
$H_T = $\met$ + \Sigma_{i}$\et$($jet$_{i})$, to demonstrate agreement with variables not directly optimized via RGS.}
\end{figure}

A random grid search (RGS) \cite{RGS} based on the energy of the two leading jets and the \met $\:$ was used to optimize the
final selection criteria to apply to the {\em clean} sample.  
RGS uses Monte-Carlo-generated
scalar top quark events to investigate the region of phase space most heavily 
populated by signal.  The RGS was run for the  mass points, 
$m_{\tilde{t}}=115$  GeV/$c^2$ and  $m_{\tilde{\chi_0}}= 20$ GeV/$c^2$, and $m_{\tilde{t}}=130$ GeV/$c^2$ and $m_{\tilde{\chi_0}}=30$ GeV/$c^2$ optimizing 
rejection of background relative to signal by  maximizing the quantity 
$N_{\text{signal}}/{ \sqrt{N_{\text{signal}}+N_{\text{background}}}}$. This was subject to the requirements of $>$ 2\% efficiency for signal. While restricting multijet backgrounds 
to account for no more than 50\% of the total background. The selection criteria as determined by RGS for each mass point was within 1-2 GeV of our final cuts, chosen to be leading jet \et$>60$ GeV, second jet \et$>50$ GeV, and \met$>60$ GeV.  Our final sample
and estimated background are reported in Table 
\ref{rgsearch_cut_table}.  Figure \ref{fig:dvsb_final} compares data and  
background for several physics distributions.  The additional contribution for a  130 GeV/$c^2$ scalar 
top quark, 30 GeV/c$^2$ neutralino signal is indicated by the cross-hatched regions on the figure.

\begin{figure}
\begin{minipage}[htb]{10.6cm} 
\epsfysize = 10.5cm 
 \centerline{\epsffile{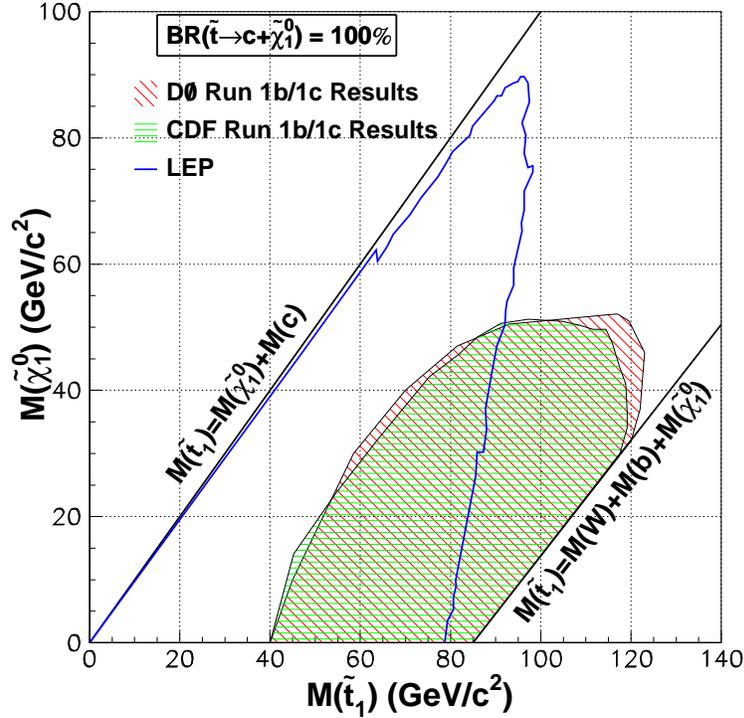}}
\end{minipage}
\caption{\label{fig:curve}Regions in the  ($m_{\tilde{t}}$,$m_{\tilde{\chi^0_{1}}}$) plane excluded at the 95\% confidence level assuming 100\% branching of $\tilde{t} \rightarrow c \chi_1^0$. Limits from LEP \cite{lep_curve} and CDF \cite{CDFstop} 
are also shown in the figure.  The dashed lines correspond to kinematic cutoffs from the masses of the $\chi_1^0$, $m_c$, $m_b$ and $m_W$.}
\end{figure}

We find the number of observed events is 
consistent with expected background.  Errors on signal efficiencies and the fraction of background
events passing final selection include the statistical uncertainties from
finite MC samples and systematics from the jet energy scale (about 7\%),
luminosity (4.3\%) and W/Z cross sections (about 6\%).  The systematic
uncertainty in simulating the trigger is dominated by
hardware trigger response (introducing a 5\% uncertainty in acceptance).   The systematic uncertainty in identifying leptons in data ranged from 2-12\% (with dependence on the detector $\eta -\phi$ position of the lepton and jet multiplicity) and the vertex confirmation procedure includes a systematic 1\% in its
efficiency. The signal acceptance of the final selection for a scalar top quark mass of $M_{\tilde{t}}$=115 GeV/$c^2$, $M_{\tilde{\chi_0}}$=40 GeV/$c^2$ was 2.7 $\pm$ 0.1\%. This null
result can be represented by a region of exclusion 
in the ($m_{\tilde{t_1}} ,  m_{\tilde{\chi}^0_1}$) plane,  
which is shown in Fig.~\ref{fig:curve} (along with results from previous experiments).  A
Bayesian method, using a flat prior for the signal cross
section and Gaussian priors for background and acceptance, sets the 95\% confidence level (CL) upper limits.    
The highest excluded scalar top quark
mass value excluded is 122 GeV/$c^2$ for a neutralino mass of 45 GeV/$c^2$.
The highest excluded neutralino mass excluded is 52 GeV/$c^2$ for a 117 GeV/$c^2$ scalar top quark mass.  
%

We thank the staffs at Fermilab and collaborating institutions,
and acknowledge support from the
Department of Energy and National Science Foundation (USA),
Commissariat  \` a L'Energie Atomique and
CNRS/Institut National de Physique Nucl\'eaire et
de Physique des Particules (France),
Ministry for Science and Technology and Ministry for Atomic
   Energy (Russia),
CAPES, CNPq and FAPERJ (Brazil),
Departments of Atomic Energy and Science and Education (India),
Colciencias (Colombia),
CONACyT (Mexico),
Ministry of Education and KOSEF (Korea),
CONICET and UBACyT (Argentina),
The Foundation for Fundamental Research on Matter (The Netherlands),
PPARC (United Kingdom),
Ministry of Education (Czech Republic),
A.P.~Sloan Foundation,
and the Research Corporation.
%


\end{document}